
\magnification 1200
\def\entier{{\rm Z}\mskip-6mu{\rm Z}}

\def\reel{{\rm I}\mskip-3.5mu{\rm R}}
\def\complexe{\lower-0.33ex\hbox{${\scriptstyle |}$}\mskip-8mu{\rm C}}

\vsize=22truecm
\hsize=15truecm
\voffset=0.5truecm
\hoffset=1truecm

\pageno=1
\baselineskip=18truept
\lineskip=2truept
\line{\hfil hep-th/yymmxxx}
\line{\hfil UdeM-LPN-TH-94-187}
\lineskiplimit=0truept
\vskip 1.5cm
\centerline{\bf Low Energy Skyrmion-Skyrmion Scattering}
\vskip 0.5cm
\centerline{by T. Gisiger and M. B. Paranjape}
\vskip 0.5cm
\centerline{\it Laboratoire de physique nucl\'eaire, Universit\'e de
Montr\'eal}
\vskip 0.5cm
\centerline{\it C.P. 6128, succ ``A", Montr\'eal, Qu\'ebec, Canada, H3C 3J7 }
\vskip 0.8cm
\centerline{\bf Abstract}
We study the scattering of two Skyrmions at low energy and large separation.
We use the method proposed by Manton for truncating the degrees of
freedom of the system from infinite to a manageable finite number. This
corresponds to identifying the manifold consisting of the union of the
low energy critical points of the potential along with the gradient flow
curves joining these together and by positing that the dynamics is
restricted here. The kinetic energy provides an induced metric on this
manifold while restricting the full potential energy to the manifold
defines a potential. The low energy dynamics is now constrained to these
finite number of degrees of freedom. For large separation of the two
Skyrmions the manifold is parametrised by the variables of the product
ansatz. We find the interaction between two Skyrmions coming
from the induced metric, which was independently found by Schroers. We find
that the static potential is actually negligible
in comparison to this interaction. Thus to lowest order, at large separation,
the dynamics reduces to geodesic motion on the manifold. We consider the
scattering to first order in the interaction using the perturbative method of
Lagrange and find that the dynamics in the no spin or charge exchange sector
reduces to the Kepler problem.
\vskip 1.0cm
\vfill\break

\baselineskip=24truept
\vskip 1.0cm
\noindent {\bf 1. Introduction and Conclusion}
\vskip 1.0cm

Scattering of solitons in a non-integrable, non-linear classical or quantum
field theory remains an intractable and difficult problem. It, however,
concerns one of the most interesting
aspects of the nature of the corresponding physics. Numerical methods have
given reasonable ideas on how the scattering proceeds but they are still
unsatisfactory for uncovering the detailed dynamics governing the scattering.

A method has been proposed by Manton$^1$ for truncating the degrees of freedom
from the original infinite number to a relevant finite number of variables for
the case of low energy scattering.
One first considers theories of the Bogomolnyi type. In these theories there is
typically a topological charge (an integer $N$) which caracterizes the soliton
sector giving in some sense the number of single
solitons in the sector. In each topological sector there exist exact static
solutions with a fixed number of arbitrary parameters, the moduli space, which
describe these solutions. The number of such parameters is equal to the
number of parameters for the one soliton solution multiplied by the topological
charge (the number of solitons). There exist asymptotic solutions which are
easily identifiable with $N$ single solitons with large mutual separations, but
for small separations the individual solitons are subject to strong
deformations and they lose their identity. As these are static solutions at
essentially arbitrary separations between the solitons, clearly there can be no
static forces between them. Otherwise the solutions would not exist; the
solitons would move towards or away from one another. Thus in each
soliton sector the sub-manifold of solutions corresponds to an equi-potential
surface, the minimal energy surface. The kinetic energy can be properly
interpreted as corresponding to a metric on the space of all configurations.
Typically we have
$$T = \int d^3\vec x \;{1\over 2}
\;g_{ij}(\phi_k)\;\dot\phi_i\dot\phi_j.\eqno(0)$$
In many instances $g_{ij}$ is just flat, but it can be non-trivial,
for example, for non-linear sigma models such as the Skyrme model. In
any case, in general, the induced metric on the sub-manifold of static
solutions is non-trivial. If one considers a time evolution with initial
conditions corresponding to a point on the sub-manifold, with initial velocity
tangent to the sub-manifold and arbitrarily small, it is intuitively evident
that
the evolution will remain on the sub-manifold and correspond simply to
geodesic motion.

It is a difficult task to prove that such a truncation of the degrees of
freedom is a sensible, mathematically rigorous, perturbative scheme. The
non-linearity of the theory means that the degrees of freedom tangential
to the equi-potential surface are coupled to all other degrees of freedom at
higher order. We are assuming that these couplings are negligible.
Gibbons and Manton$^2$ applied this program with remarkable success to the case
of magnetic monopoles in the BPS limit and it has also been applied to many
other situations, including vortex scattering, Kaluza Klein monopoles, black
holes, lumps in $CP^N$ models, etc. (We refer the reader to the article of
Samols$^3$ for detailed references).

The generalization to the more common situation where the set of static
solutions do not include solitons at essentially arbitrary separations
can also be developed. Here the forces between the solitons do not
exactly cancel, but these are assumed to be weak. The moduli space of the
minimum energy critical point is of a smaller dimension than before and does
not ever represent many isolated single solitons. Thus truncation of the
dynamics to the sub-manifold of such configurations is inadequate to describe
the scattering of the solitons. It is, however, evident that the moduli space
of the asymptotic critical point of $N$ infinitely separated single solitons
will in fact be $N\times k$ dimensional, where $k$ is the dimension of the
moduli space for one soliton. It is argued by Manton, that the low energy
scattering will be restricted to or will lie close to the part of the
configuration
space corresponding to the union of the gradient flow curves which start at the
asymptotic critical point and reach other critical points. They will reach the
minimal critical point or, in
fact, other critical points, which may actually have energy more or less than
the starting point.
We find this idea also intuitively reasonable. As long as the gradients
involved are not too steep very little radiation should be incurred. For the
Skyrme model, numerically it is known$^1$ in the two soliton sector that the
difference in energy between
the ``deuteron'' bound state and the asymptotic, infinitely separated,
configuration is rather small, thus the gradients involved should not be large.

In this paper we identify the sub-manifold corresponding to the gradient flow
curves for the Skyrme model, in the two Skyrmion sector for large separation
between the Skyrmions, as the set of configurations parametrized by the product
ansatz (undeformed). The force between two Skyrmions at large distance is
governed by their
far field behaviour. For the product ansatz this gives rise to an interaction
which falls off like $1/d^3$, where $d$ is the separation. Localized
deformations of each Skyrmion will not affect the behaviour of the interaction.
Modification of the far field behaviour of each Skyrmion will only weaken their
mutual interactions if the self energy of the modified Skyrmions is to be kept
finite. Modifications which sharpen the far field fall off are permitted while
those which weaken the fall off cause the self energy to diverge. We are not
concerned here with intermediate range modifications and interactions.
Therefore it is clear that the product ansatz with undeformed Skyrmions at
large separation gives a good parametrization of the manifold of gradient flow
curves up to order $1/d^3$. This fact is confirmed by numerical calculations
where
it is observed that the undeformed product ansatz is surprisingly accurate
even down to separations of the order of the size of a Skyrmion$^4$.

We compute the induced metric on this manifold coming from the kinetic energy.
We actually
find that the metric dominates the interactions between the Skyrmions,
the static potential is relatively negligible. To first order in the inverse
separation the free metric is modified by a term which behaves like $1/d$,
which is much greater than the potential which behaves like $1/d^3$. The
characteristic scale is set by $f_\pi$, by $e$ and the details of the Skyrmion
configuration. It is reasonable to expect that it is given by the size of the
Skyrmion. We
remark that this scale has nothing to do with the scale given by a light
massive pion. We have considered a massless pion, adding a small mass does not
drastically modify the Skyrmion size. At
large distance of course the pion mass cuts off all interactions with the
usual Yukawa exponential. Thus our results are valid for the separation in the
range that is large compared with the Skyrmion size but small compared with the
Compton wavelength of the pion.

The scattering, to first order, is again described by geodesic motion
on the manifold parametrized by the product ansatz variables. Even though this
is a great simplification from an infinite number of degrees of freedom to just
12 collective coordinates, the problem remains intractable. We calculate 2
conserved ``angular'' momenta coming from invariance under right iso-rotation
coupled with spatial rotation and single left iso-rotation. Then we use the
Lagrange perturbative method to obtain approximate equations of motion which
correspond to a systematic expansion in the inverse separation. It is actually
quite reasonable to make this further approximation since
we have
already dropped terms of second order in the induced metric. The equations are
still rather complicated and we expect a rich structure in the scattering cross
section. We show that the system reduces simply to the Kepler problem for
Skyrmions
synchronously rotating in a direction orthogonal to the scattering plane.

\vskip 2.0cm
\noindent {\bf 2. The Skyrme model}
\vskip 1.0cm

The Skyrme model is described by the Lagrangean density,
$${\cal L}_{sk} =
-{f_\pi^2 \over 4}\; tr(U^\dagger \partial_\mu U U^\dagger \partial^\mu U)
+ {1\over 32 e^2}\;tr( [U^\dagger \partial_\mu U,U^\dagger \partial_\nu U]^2)
\eqno(1)$$
where $U(x)$ is a unitary matrix valued field. We take
$$U(x) \in SU(2).$$
The Skyrme Lagrangean corresponds to the first two terms of a systematic
expansion in derivatives of the effective Lagrangean describing low energy
interaction of pions. It should be derivable from QCD hence $f_\pi$ and $e$ are
in principle calculable parameters. These calculations are actually unfeasable
and we take $f_\pi$ and $e$ from phenomenological fits.
What is even more surprising is that the Skyrme model includes
the baryons as well. These arise as topological soliton solutions of the
equations of motion. The original proposal of this by Skyrme$^5$ in the 60's
was put on solid footing by Witten$^6$ in the 80's. For further references and
details see the review article by Wambach and Walhout$^7$.

The topological solitons, called Skyrmions, correspond to non-trivial mappings
of $\reel^3$ plus the point at infinity into $SU(2)$:
$$U(x): \reel^3 + \infty \to SU(2) = S^3.\eqno(2)$$
But
$$\reel^3 + \infty = S^3\eqno(3)$$
thus the homotopy classes of mappings
$$U(x): S^3 \to S^3\eqno(4)$$
which define
$$\Pi_3(S^3) = \entier\eqno(5)$$
characterize the space of configurations.

The topological charge of each sector is given by
$$ N = {1 \over 24 \pi^2} \int{d^3 \vec x\, \epsilon^{ijk} \;
tr(U^\dagger \partial_i U U^\dagger \partial_j U U^\dagger \partial_k U)}
\eqno(6)$$
which is integer and is identified with the baryon number. Thus for the
scattering of two
Skyrmions, we are looking at the sector of baryon number equal to 2.
In this sector the minimum energy configuration should correspond to the bound
state of two Skyrmions, which must represent the deuteron. The asymptotic
critical point corresponds to two infinitely separated Skyrmions.
There exist, known, non-minimal critical points, corresponding to a spherically
symmetric configuration, the di-baryon solution$^8$. The energy of this
configuration is about three times the energy of a single Skyrmion. There
are also, possibly, other non-minimal critical points with energy less than two
infinitely separated Skyrmions$^9$. The scattering of two
Skyrmions will take place
on the union of the paths of steepest descent which connect the various
critical points.
\vfill
\eject
\vskip 2.0cm
\noindent {\bf 3. Lagrangean of the Skyrmion-Skyrmion system}
\vskip 1.0cm

With the definitions
$${\cal L}^a_\mu(U)=-{i\over 2}\; tr[\tau^a \partial_\mu U U^\dagger]\eqno(7)$$
$${\cal R}^a_\mu(U)=-{i\over 2}\; tr[\tau^a U^\dagger \partial_\mu U]\eqno(8)$$
$$D_{ab}(U) = {1\over 2}\;tr[\tau^a U \tau^b U^\dagger]\eqno(9)$$
the Skyrme Lagrangean density becomes
$$\eqalign{
{\cal L}_{sk}&={f_\pi^2\over 2} {\cal L}(U)_\mu\cdot{\cal L}(U)^\mu\cr
&- {1\over{4 e^2}}\biggl [{\cal L}(U)_\mu\cdot{\cal L}(U)^\mu\;
{\cal L}(U)_\nu\cdot{\cal L}(U)^\nu -
{\cal L}(U)_\mu\cdot{\cal L}(U)^\nu\;{\cal L}(U)_\nu\cdot{\cal
L}(U)^\mu\biggr]\cr }\eqno(10)$$
where the $\cdot$ implies a contraction in isospace.
We can separate this into a kinetic energy ${\cal T}$ which is the part
quadratic in time derivatives and a potential energy ${\cal V}$ without any
time derivatives:
$${\cal T}={f_\pi^2\over 2} {\cal L}_0\cdot{\cal L}_0 -
{1\over{2 e^2}}\biggl [{\cal L}_0\cdot{\cal L}_i\;
{\cal L}_0\cdot{\cal L}_i -
{\cal L}_0\cdot{\cal L}_0\;{\cal L}_i\cdot{\cal L}_i\biggr ]\eqno(11)$$
$${\cal V}=-{f_\pi^2\over 2} {\cal L}_i\cdot{\cal L}_i -
{1\over{4 e^2}}\biggl [{\cal L}_i\cdot{\cal L}_i\;
{\cal L}_j\cdot{\cal L}_j -
{\cal L}_i\cdot{\cal L}_j\;{\cal L}_j\cdot{\cal L}_i\biggr ].\eqno(12)$$

We consider the scattering only for large separation, thus we do not
have to know the structure of the complicated
region where the two Skyrmions interact strongly and consequently are much
deformed. In the region of large separation the product ansatz corresponds to
$$
\eqalign{
U(\vec x) &= U_1(\vec x - \vec R_1)\; U_2(\vec x - \vec R_2)\cr
          &= A U(\vec x - \vec R_1)A^\dagger\;B U(\vec x - \vec R_2)B^\dagger
}\eqno(13)$$
where $U(\vec x - \vec R_1)$ and $U(\vec x - \vec R_2)$ correspond to the field
of a single Skyrmion solution centered at
$\vec R_1$ and $\vec R_2$ respectively. This particular ansatz gives a 12
dimensional configuration space.
It can be seen that
$${\cal L}^a_\mu(U_1 U_2) = {\cal L}^a_\mu(U_1) + D_{ab}(U_1)\;{\cal
L}^b_\mu(U_2)\eqno(14)$$
so ${\cal T}$ and ${\cal V}$ each separate into three parts: one depending only
on $U_1$, another depending only on $U_2$ and a third function of both $U_1$
and
$U_2$ which describes the interaction between Skyrmions. This last term has
already been investigated for the potential part
${\cal V}$, and gives a contribution of the order $1/d^3$ where $d=||\vec
R_1-\vec R_2||$ (For details see Ref. 9 and references therein). We shall
concentrate here on the interaction part of the kinetic energy ${\cal T}$ given
by:
$$\eqalign{
{\cal T}_{int} = f_{\pi}^2\; {\cal L}_0^1\cdot D\cdot {\cal L}_0^2 & \cr
+{1\over 2 e^2}
\Biggl[ \bigl({\cal L}_0^1\cdot {\cal L}_0^1\; &{\cal L}_i^2\cdot {\cal L}_i^2
+1\leftrightarrow 2\bigr)
+2\bigl({\cal L}_0^1\cdot{\cal L}_0^1+1\leftrightarrow 2\bigr) {\cal
L}_i^1\cdot D\cdot{\cal L}_i^2\cr
+2\;\bigl( {\cal L}_i^1\cdot &{\cal L}_i^2+1\leftrightarrow 2\bigr) {\cal
L}_0^1\cdot D\cdot{\cal L}_0^2
+4\;{\cal L}_0^1\cdot D\cdot{\cal L}_0^2\;
{\cal L}_i^1\cdot D\cdot{\cal L}_i^2\cr
-\;\bigl({\cal L}_0^1\cdot &D \cdot {\cal L}_i^2\bigr)^2
 -\bigl({\cal L}_i^1\cdot D\cdot {\cal L}_0^2\bigr)^2
-2\;{\cal L}_0^1\cdot {\cal L}_i^1\;{\cal L}_0^2\cdot {\cal L}_i^2\cr
-\;2\;{\cal L}_0^1\cdot &D \cdot{\cal L}_i^2\;
{\cal L}_i^1\cdot D\cdot{\cal L}_0^2\cr
&-2\;\bigl({\cal L}_0^1\cdot {\cal L}_i^1+1\leftrightarrow 2\bigr)
\bigl( {\cal L}_0^1\cdot D\cdot {\cal L}_i^2 +
       {\cal L}_i^1\cdot D\cdot {\cal L}_0^2\bigr)
\Biggr]\cr}\eqno(15)$$
where ${\cal L}_\mu^1\equiv {\cal L}_\mu^a(U_1)$, ${\cal L}_\mu^2\equiv {\cal
L}_\mu^a(U_2)$ and $D \equiv D_{ab}(U_1)$.
Since the time dependance of $U_1$ is contained in $A$ and
$\vec R_1$, we find that
$${\cal L}^a_0(U_1)=\biggl(\delta^{ab}-D_{ab}\bigl(AU(\vec x-\vec R_1)
A^\dagger\bigr) \biggr) \;{\cal L}^a_0(A)
-D_{ab}(A)\;\dot{\vec R^i_1}\;{\cal L}^b_i\bigr(U(\vec x-\vec
R_1)\bigl)\eqno(16)$$
and correspondingly for $U_2$, $B$ and $\vec R_2$.

\noindent
It can be seen that $(\delta^{ab}-D_{ab}(AU(\vec x-\vec R_1) A^\dagger))$ is of
order $1/r^2$ while ${\cal L}^a_i(U(\vec x-\vec R_1))$ is of order $1/r^3$ at
large distance due to the behaviour of $F$ which falls off like $1/r^2$. This
actually implies below that the term containing ${\cal L}^b_i$ above is not
important to the term of leading order in the kinetic energy.

We now outline the computation of the kinetic energy $T_{int}$ obtained after
integrating ${\cal T}_{int}$ over all space. The procedure is similar to the
one used in Ref. 9 to calculate the interaction from the potential. We divide
space  into three regions I, II an III. II is a spherical region centered on
the first Skyrmion, of a radius much smaller than $d$ but large enough so that
outside region II the asymptotic behaviour of $F$ is valid. Region III is the
same for the second Skyrmion. Region I is of course the remaining space and the
field coming from both Skyrmions behave asymptotically here. We find the
leading contribution to $T_{int}$ behaves as $1/d$ and comes from the first
term in ${\cal T}_{int}$ evaluated in region I. The contributions coming from
the Skyrme term are of higher order due to the behaviour of ${\cal L}^b_i$. The
leading contributions from region II and III are of order $1/d^2$ and hence
negligible. The interaction over region II is actually computed by extending
the integrand to the whole of space. This is justified since it only modifies
the contribution at higher order in $1/d$.

This simplifies the evaluation of the integral and we find
$$\eqalign{
T_{int} = &\int\!d^3 x f_{\pi}^2 {\cal L}_0^a(U_1)\,D_{ab}(U_1)\,
{\cal L}^b_0(U_2) +  O(1/d^2)\cr
&={\Delta\over d} \epsilon^{iac}\epsilon^{jbd}\; {\cal R}^a(A) \,{\cal
R}^d(B)\,
\bigl(\delta^{ij}-\hat{d}^i \hat{d}^j\bigr) \,D_{ab}(A^\dagger B)
+O(1/d^2)}\eqno(17)$$
where $\Delta=2 \pi \kappa^2 f_\pi^2$, $F(r)\sim \kappa/r^2$ at large
$r$, ${\cal R}^a (A)\equiv {\cal R}^a_0(A)$ and $\hat d = \vec d/d$. We have
used the relation ${\cal R}^a(A)=D_{ab}(A)\;{\cal L}^b(A)$.
The free part of the kinetic energy is well known$^7$ and we can finally write
the Lagrangean
$L$ for the $N=2$ sector of the Skyrme model to leading order of the dynamics
of the variables of the product ansatz:
$$\eqalign{
L =& -2 M + {1\over 4} M \dot{\vec d^{\;2}} + 2 \Lambda \bigl({\cal L}^a(A)\,
{\cal L}^a(A) + {\cal L}^a(B)\,{\cal L}^a(B)\bigr)\cr
& + {\Delta\over d} \epsilon^{iac}\epsilon^{jbd}\;{\cal R}^a(A)\,{\cal
R}^d(B)\;
\bigl(\delta^{ij}-\hat{d}^i \hat{d}^j\bigr)\, D_{ab}(A^\dagger B) +O(1/d^2)}
\eqno(18)$$
where
$$\eqalign{M =\;& 4\pi \int_0^\infty r^2dr\cr
\times &{ \biggl\{ {1\over 8} f_\pi^2 \biggl[\biggl({\partial F\over \partial
r}\biggr)^2\!\!+ 2\,{\sin^2 F\over r^2}\biggr]+{1\over 2 e^2}{\sin^2 F\over
r^2}
\biggl [{\sin^2 F\over r^2} + 2\biggl({\partial F\over \partial
r}\biggr)^2\biggr] \biggr\} }\cr}\eqno(19)$$
is the mass of a Skyrmion and
$$\Lambda = (ef_\pi )^3\int{r^2 dr \sin^2 F\biggl[ 1+ {4\over (ef_\pi)^2 }
\biggl(F'^2+ {\sin^2 F\over r^2}\biggr)\biggr] }\eqno(20)$$
is its inertia momentum. We have replaced $\vec R_1-\vec R_2$ by $\vec d$
placing us in the center of mass reference frame and reducing the number of
degrees of freedom of the system to 9. Equation (18) is the main result of our
calculation.

The metric can easily be obtained from this expression by choosing local
coordinates on the product ansatz manifold and extracting the quadratic form
relating their time derivatives. With the potential part absent from the
Lagrangean of the system (to first approximation) the solution of the problem
now resides in finding the geodesics of the metric on the product ansatz
manifold.

The result of equation (18) was also independently obtained by
Schroers$^{10}$.  He found a leading contribution which behaves as $1/d$ and
even calculated sub-leading spin-orbit coupling terms.
The only other comparable calculation to our knowledge has been done by Walhout
and Wambach$^7$ for the case of massive pions. The limit as
$m_\pi\rightarrow 0$ of their
expression, however, does not leave a term which behaves as $1/d$ and hence
does not reproduce our result. We believe
that this contribution should come also from their evaluation of the integral
giving the induced kinetic energy in the far field region (region I) and then
recovering our result as $m_\pi\rightarrow 0$. Such a contribution would also
be
proportional to $(e^{-m_\pi d})^2$, what they call ``two pion exchange''. We
believe that they have not computed the contribution from this region.

We also add that our term is leading order in an expansion in inverse
separation with respect to a scale which has nothing to do with the pion mass.
This means that there are two length scales for nucleon-nucleon interaction
predicted by the Skyrme model. It would be interesting to see if this can be
phenomenologically or experimentally justified.

\vskip 2.0cm
\noindent {\bf 4. Approximate Euler-Lagrange equations}
\vskip 1.0cm

Unfortunately the two body Lagrangean has a very complicated
structure. Simply finding the expression of the Euler-Lagrange equations is a
long and tedious undertaking (let alone finding solutions to these
equations). We do not record the equations here.
The Lagrangean possesses numerous symetries and corresponding conserved
quantities. There are two interesting symetries apart from translational
invariance giving conserved charges.

{}From left iso-rotation
$$\eqalign{A &\rightarrow CA\cr
           B &\rightarrow CB\cr}\eqno(21)$$
where $C$ is a constant $SU(2)$ matrix we find the conserved ``total isospin'':
$$\eqalign{J_L^k = 4 \Lambda \bigl[ {\cal L}^k(A) + {\cal L}^k(B) \bigr]&\cr
+{\Delta\over d}\Bigl[\epsilon^{kne} D_{ni}(A) &(\delta^{ij}-\hat d^i\hat d^j)
\epsilon^{jbd} D_{eb}(B) {\cal R}^d(B)\cr
 &+ \epsilon^{kne} D_{nj}(B) D_{ea}(A) \epsilon^{iac} {\cal R}^c(B)
(\delta^{ij}-\hat d^i\hat d^j) \Bigr].\cr}\eqno(22)$$
The Lagrangean is also invariant under right iso-spin rotation coupled with a
spatial rotation
$$\eqalign{ A &\rightarrow AC\cr
            B &\rightarrow BC\cr
            d^a &\rightarrow D_{ab}(C^\dagger) d^b\cr}\eqno(23)$$
where $C$ is a constant $SU(2)$ matrix and $D_{ab}(C)$ is its
representative for spin 1. This gives the further conserved ``total angular
momentum'':
$$\eqalign{J_R^k = M \bigl(\vec d\times \dot{\vec d}\bigr)^k +
4 \Lambda \bigl[ {\cal R}^k(A) + &{\cal R}^k(B) \bigr]\cr
+{\Delta\over d} \Bigl[ \epsilon^{iak} \epsilon^{jbd} {\cal R}^d(B)
(\delta^{ij}-&\hat d^i\hat d^j) D_{ab}(A^\dagger B)\cr
+&\epsilon^{iac} \epsilon^{jbk} {\cal R}^c(A) (\delta^{ij}-\hat d^i\hat d^j)
D_{ab}(A^\dagger B)\Bigr].\cr}\eqno(24)$$
These conserved quantities do not help us greatly in solving the equations for
the geodesics.
In fact
it should be noted that the peculiar form of the Lagrangean (not unlike one
describing a pair of coupled rigid bodies floating in free space) does not even
allow the usual separation of the rotational movement into a global and
relative part.
In order to go further we must resort to a supplementary approximation.

We use the perturbation method of Lagrange$^{11}$ familiar in celestial
mechanics.
This approximation scheme neglects the changes
induced on the free canonical momenta by the interaction term. To begin with
we need
the free Poisson brackets for the degrees of freedom of the system.
We use the variables $(\vec d, \dot{\vec d}, {\cal L}^a(A), {\cal L}^a(B),
{\cal R}^a(A), {\cal R}^a(B)).$
${\cal L}^a$ is proportional to the isospin of a \break Skyrmion and
${\cal R}^a$ to its spin.
We can easily compute the Poisson brackets for the free theory (the limit $d$
goes to infinity of $L$):
$$\{ d^i,\Pi^j\} = \delta^{ij}$$
$$\{{\cal L}^a(A),{\cal L}^b(A)\} = -{1\over 2\Lambda} \epsilon^{abc} {\cal
L}^c(A)$$
$$\eqalign{\{{\cal R}^a(A),{\cal R}^b(A)\} = {1\over 2\Lambda}
&\epsilon^{abc} {\cal R}^c(A)\cr
{}&\cr
\{{\cal L}^a(A),{\cal R}^b(A)\} &= 0\cr}\eqno(25)$$
$$\{{\cal L}^a(A),D_{bc}(A)\} = -{1\over 2\Lambda} \epsilon^{abd}D_{dc}(A)$$
$$\{{\cal R}^a(A),D_{bc}(A)\} = {1\over 2\Lambda} \epsilon^{acd}D_{db}(A)$$
where $\Pi^i$ is the conjugate momenta to $d^i$. Because of the symmetric
nature
of the free Hamiltonian, the same brackets are true if we replace $A$ by $B$
everywhere, and all the mixed brackets between $A$ and $B$ are zero.
The free system in the center of mass reference frame possesses 5 vectorial
conserved quantities: $\dot d^i$, ${\cal L}^a(A)$, ${\cal L}^a(B)$, ${\cal
R}^a(A)$ and ${\cal R}^a(B)$. Let $C^i$ be one of those quantities. Its time
derivative is given by
$${d\over dt} C^i = \{C^i,H\}$$
where $H=H_{free} + H_{int}$.
The approximation comes by using the free Poisson brackets to
compute $\{C^i, H_{int}\}_{_0}$. This is of course only correct to first order.
Since $C^i$ is conserved in the free system,
$\{C^i,H_{free}\}_{_0}=0$ and
$${d\over dt} C^i \simeq \{C^i, H_{int}\}_{_0}.$$
With the free Poisson
brackets we obtain without difficulty the following coupled
system of equations:
\vfill
\eject
$$\eqalignno{
\ddot d^k + {2\Delta\over M d^2}\biggr[\delta^{ij}\hat d^k + \delta^{jk}\hat
d^i +&\delta^{ik}\hat d^j - 3\hat d^i \hat d^j \hat d^k\biggl] \epsilon^{iac}
\epsilon^{jbd}
{\cal R}^c(A) {\cal R}^d(B) D_{ab}(A^\dagger B) = 0\cr
{d\over dt}{\cal L}^k(A) =  {\Delta\over 2 M d} \epsilon^{iac} \epsilon^{jbd}
{\cal R}&^c(A) {\cal R}^d(B) \bigl(\delta^{ij}-\hat d^i\hat d^j\bigr)
\epsilon^{kef} D_{fa}(A)D_{eb}(B)\cr
{d\over dt}{\cal L}^k(B) =  {\Delta\over 2 M d} \epsilon^{iac} \epsilon^{jbd}
{\cal R}&^c(A) {\cal R}^d(B) \bigl(\delta^{ij}-\hat d^i\hat d^j\bigr)
\epsilon^{kef} D_{ae}(A^\dagger )D_{fb}(B)\cr
{d\over dt}{\cal R}^k(A) = - {\Delta\over 2 M d}\epsilon^{iac} \epsilon^{jbd}
&{\cal R}^d(B)\bigl(\delta^{ij}-\hat d^i\hat d^j\bigr) &(26)\cr
\times & \biggl[\epsilon^{kcf} {\cal R}^f(A) D_{ab}(A^\dagger B) +
\epsilon^{kaf} D_{fb}(A^\dagger B) {\cal R}^c(A)\biggr]\cr
{d\over dt}{\cal R}^k(B) = - {\Delta\over 2 M d}\epsilon^{iac} \epsilon^{jbd}
&{\cal R}^c(A)\bigl(\delta^{ij}-\hat d^i\hat d^j\bigr)\cr
\times & \biggl[\epsilon^{kdf} {\cal R}^f(B) D_{ab}(A^\dagger B) +
\epsilon^{kbf} D_{af}(A^\dagger B) {\cal R}^d(B)\biggr].\cr
}$$
Our approximation is reliable as long as the separation
$d$ between the particles is large enough for the conjugate momenta to stay
close to their free values. As we have already worked with the undeformed
product ansatz approximation, and neglected the potential, which are both valid
for large $d$, we feel confident
that we have not lost any meaningful information by making this further
approximation. If $d$ is kept large we should then find geodesics similar
(qualitatively at least) to those given by the exact equations of motion.
The system of equations (28) is still quite complicated but some
specialized solutions are easy
to obtain and give simple trajectories as we shall now see.

\vskip 2.0cm
\noindent {\bf 5. Skyrmion-Skyrmion scattering}
\vskip 1.0cm

The simplest geodesic is obtained by taking $B=A$ and replacing it in the
equations (28). We then get the following system of equations:
$$\ddot d^k + {2\Delta\over M d^2}\biggl[\hat d^k\Bigl({\cal R}^a(A)
{\cal R}^a(A) +
3 \bigl(\hat d^a {\cal R}^a(A)\bigr)^2\Bigr) - 2 {\cal R}^k(A) \hat d^a
{\cal R}^a(A)\biggr]=0$$
$${d\over dt}{\cal L}^k(A) = 0\eqno(27)$$
$${d\over dt}{\cal R}^k(A) = - {\Delta\over 2 M d} {\cal R}^a(A) \hat d^a
\epsilon^{kbc} \hat d^b {\cal R}^c(A)$$
We see that
if $A$ is chosen at $t=0$ such that $\hat d$ is parallel or perpendicular to
${\cal R}^k(A)$ then the last two equations become $\dot{{\cal L}^k}=0$ and
$\dot{{\cal R}^k}=0$ and are satisfied if ${\cal L}^k(A)$
and ${\cal R}^k(A)$ are constants. This means that $A$ corresponds to an
iso-rotation about a fixed axis with constant angular velocity.

Let us begin by choosing ${\cal R}^k(A)$ parallel to $\hat d$. This does not
actually lead to a scattering because $\hat d$ does not have the liberty to
change direction with time (it has to stay parallel to ${\cal R}^k(A)$ which is
a constant). Substituting the condition ${\cal R}^k=\alpha\; \hat d^k=
\hbox{constant}$ in the equation for $\vec d$ we find an equation for a
particle constrained on a line with an attractive Coulomb potential. The
trajectory will necessarily lead to $d$ small and hence to regions where we can
no longer trust our equations or their predictions.

The case where ${\cal R}^k(A)$ is perpendicular to $\hat d$ is more
interesting. Again the last two equations are satisfied with $A$ corresponding
to a steady iso-rotation, however $\hat d$ now has the freedom to move in a
plane according to the equation:
$$\ddot d^k + {2 \Delta\over M d^2} {\cal R}^a(A){\cal R}^a(A) \hat d^k = 0.
\eqno(28)$$
This is the equation of a Kepler system: the two Skyrmions scatter in a plane
while keeping their isospins and spins constant and
perpendicular to the plane of the orbit.

These geodesics are the simplest to compute algrebraically but are almost all
that we can
compute by hand. Indeed if we take ${\cal R}^a$ at $t=0$ not exactly
parallel or
perpendicular to $d^a$, complicated precessions appear. It seems that
the typical motions of the Skyrmions described by equations (28) are very
complex and allow for large nutations, precessions and spin flips. Quantum
mechanically these would correspond to neutral and charged pion exchange which
we expect from a dynamical
model representing nucleon-nucleon interactions.
\vskip 2.0cm
\centerline{\bf Aknowlegements}
\vskip 1.0cm
We thank D. Caenepeel, O. Hernandez, R. Mackenzie, M.Temple-Raston and
P. Winternitz for useful discussions. This work supported in part by NSERC of
Canada and FCAR of Qu\'ebec.
\vfill\break

\vskip 1.0cm
\centerline{\bf References}
\vskip 2.0cm

\noindent{[1] N.S. Manton, {\it Phys. Rev. Lett.} {\bf 60} 1916 (1988)}
\vskip 0.3cm
\noindent{[2] G.W. Gibbons, N. S. Manton, {\it Nucl. Phys.} {\bf B274} 183
(1986)}
\vskip 0.3cm
\noindent{[3] T.M. Samols, {\it Comm. Math. Phys.} {\underbar{145}}, 149
(1992)}
\vskip 0.3cm
\noindent{[4] T.S. Walhout, J. Wambach, {\it Phys. Rev. Lett.}
{\underbar {67}}, 314 (1991)}
\vskip 0.3cm
\noindent{[5] T.H.R. Skyrme, {\it Proc. Roy. Soc. Lon.}
{\underbar{260}}, 127 (1961)}
\vskip 0.3cm
\noindent{[6] E. Witten, {\it Nucl. Phys.} {\bf B223}, 422, 433
(1983)}
\vskip 0.3cm
\noindent{[7] T.S. Walhout, J. Wambach, {\it Int. Jour. Mod. Phys.} {\bf E1}
665
(1992)}
\vskip 0.3cm
\noindent{[8] M. Kutschera, C.J. Pethick, {\it Nucl. Phys.} {\bf A440} 670
(1985)}
\vskip 0.3cm
\noindent{[9] K. Isler, J. LeTourneux, M.B. Paranjape, {\it Phys. Rev.} {\bf
D43}, 1366 (1991)}
\vskip 0.3cm
\noindent{[10] B. Schroers, Durham preprint, DTP 93-29, hep-th/9308017}
\vskip 0.3cm
\noindent{[11] \underbar{Classical Mechanics}, H. Goldstein, Second edition,
Addison-Wesley publishing company 1980, 505}
\end